\documentclass{ws-procs9x6}

\begin{document}

\title{Quantum fluctuations of topological ${\mathbb S}^3$-kinks}

\author{A. Alonso Izquierdo$^{(a)}$,
M. A. Gonzalez Leon$^{(a)}$, \\ J. Mateos Guilarte$^{(b)}$, M. J.
Senosiain$^{(c)}$}

\address{ $^{(a)}$ Departamento de Matematica
Aplicada and IUFFyM \\ $^{(b)}$ Departamento de Fisica Fundamental and IUFFyM \\
$^{(c)}$ Departamento de Matematicas, Universidad de Salamanca, SPAIN}

\begin{abstract}
The kink Casimir effect in the massive non-linear $S^3$-sigma model is analyzed.
\end{abstract}


\bodymatter

\section{Introduction}
Quantum fluctuations around background kink fields are sophisticated
cousins of vacuum fluctuations. Van Nieuwenhuizen et al.\cite{RvNW0}{}
reported on the state of the art in this topic in QFEXT03 for susy
solitons as pointed out by Milton\cite{Mil}{}. More
recently, new results have been achieved by (almost) the same Stony
Brook/Wien group in the analysis of the quantum fluctuations of susy
solitons of non-linear sigma models \cite{MRvNW,RSvN}{}. Almost in parallel, we developed a similar program \cite{AMAJ,AMAJ1,AMAJM,AMAJW}{} for the kinks of the
massive non-linear ${\mathbb S}^2$-sigma model in a purely bosonic
framework. Our goal in this work is to describe the quantum fluctuations of the ${\mathbb
S}^3$-kinks. The bosonic sector of the non-linear version of the Gell-Mann/Levy
$\sigma$-model \cite{GML}{} is precisely the system that we are
going to address. Being non renormalizable in (3+1)-dimensions, it
was conceived as an effective theory describing the low energy
interactions of nucleons and pions. In (1+1)-dimensions, however,
the pion dynamics can be re-interpreted as the dynamics of a linear
chain of ${\mathbb O}(4)$ spin fields, which was renormalized by
Brezin et al\cite{BZJ}{}. We just merely add
quadratic terms in the fields to escape from infrared divergences.

\section{Massive non-linear ${\mathbb S}^3$-sigma model and topological kinks}

Let us consider $\phi_a(t,x) \, , \, a=1,2,3,4$, four scalar fields
in the $(1+1)$-dimensional Minkowski space-time ${\mathbb R}^{1,1}$.
The action of the massive non-linear ${\mathbb S}^3$-sigma model
looks very simple
\begin{equation}
S[\phi_1,\phi_2,\phi_3,\phi_4]=\int \, dtdx \,
\left\{\frac{1}{2}g^{\mu\nu}\sum_{a=1}^4\frac{\partial\phi_a}{\partial
x^\mu}\frac{\partial\phi_a}{\partial x^\nu}-\frac{1}{2}
\sum_{a=1}^4\, \alpha_a^2 \, \phi_a^2(x^\mu) \right\}\,,
\label{ncact}
\end{equation}
where $\alpha_1^2>\alpha_2^2>\alpha_3^2>\alpha_4^2$,
but the fields are constrained to live in the ${\mathbb S}^3$-sphere,  $\phi_1^2+\phi_2^2+\phi_3^2+\phi_4^2=R^2$ forming the infinite dimensional space: ${\rm Maps}({\mathbb
R}^{1,1},{\mathbb S}^3)$. We take $g_{\mu\nu}={\rm
diag}(1,-1,-1,1)$ and the natural system of
units $\hbar=c=1$. We select $\sigma_2^2=\frac{\alpha_2^2-\alpha_4^2}{\alpha_1^2-\alpha_4^2}=\frac{\gamma_2^2}{\lambda^2}$,
$\sigma_3^2=\frac{\alpha_3^2-\alpha_4^2}{\alpha_1^2-\alpha_4^2}=\frac{\gamma_3^2}{\lambda^2}$,
such that $0 < \sigma_3^2 < \sigma_2^2 < \sigma_1^2=1$, and define
non-dimensional coordinates: $x^\mu \rightarrow
\frac{x^\mu}{\lambda}$.

The extremely non-linear dynamics implied by (\ref{ncact}) plus the constraint
is unveiled if one solves $\phi_4$ in favor of $\phi_1,
\phi_2, \phi_3$ in the action and introduce the power expansion of the non-polynomial term. This process shows that: (a) There are an infinite number of vertices determining
the interactions between the three pseudo-Nambu-Goldstone bosons.
(b) $\frac{1}{R^2}$ is the coupling constant. Vertices with
different numbers of legs belong to different orders of perturbation
theory: $\frac{1}{R^{2n-2}}$ arises as a factor in the vertices with
$2n$ legs. (c) In $(1+1)$-dimensions massless bosons are discarded
due to the infrared asymptotics. We consider the
situation when the three masses are different. The one-loop self-energy graphs of $\phi_1$, $\phi_2$ and $\phi_3$:
$\Sigma_2(\sigma_2,\sigma_3)=\sigma_2^2\Sigma_1(\sigma_2,\sigma_3)$,
$\Sigma_3(\sigma_2,\sigma_3)=\sigma_3^2\Sigma_1(\sigma_2,\sigma_3)$
are divergent because $\Sigma_1(\sigma_2,\sigma_3)=\frac{2i}{R^2}\left(
I(1)+I(\sigma_2^2)+I(\sigma_3^2)\right)$ with $I(c^2)=\int
\frac{dk}{4\pi} \frac{1}{\sqrt{k^2+c^2}}$.
To tame these infinities the one-loop mass renormalization
counter-terms
\[
{\displaystyle {\cal
L}_{CT}=-\frac{1}{R^2}\cdot\Sigma_1(\sigma_2,\sigma_3)
\left(\phi_1^2(x^\mu)+\sigma_2^2\phi_2^2(x^\mu)+\sigma_3^2\phi_3^2(x^\mu)\right)}
\]
must be added to the bare Lagrangian. Searching only for
semi-classical effects we do not need to care about other divergent
graphs.

The classical minima of the action are the static and homogeneous
configurations that annihilate the integrand in (\ref{ncact}), i.e., the North and South Poles of ${\mathbb S}^3$. There is the
possibility of the existence of topological kinks and to search for
them it is convenient to use polar hyper-spherical coordinates: $\phi_1 =R \sin\psi \sin \theta  \cos \varphi$, $\phi_2 = R \sin\psi \sin \theta  \sin \varphi$,
$\phi_3 = R \sin\psi \, \cos \theta$, $\phi_4 = R
\cos \psi$, $\psi \in  [0,\pi )$,$\theta \in
[0,\pi)$, $\varphi \in [0,2\pi)$. There are three types of these kinks: (1) in the meridians on the $\phi_3-\phi_4$ plane, $\theta=0$ or
$\pi$, the non-trivial field equation is: $\frac{\partial^2 \psi }{\partial t^2}-\frac{\partial^2 \psi }{\partial x^2}+\frac{\sigma_3^2}{2} \, {\rm sin}2\psi = 0$ and the kink solutions, that we shall denote generically as $K_1$, can be written in the form $\psi_{K_1}(t,x)\, =\, 2\, {\rm arctan} \, e^{\pm \sigma_3
\overline{x}}$ where $\overline{x} =\frac{x-x_0-vt}{\sqrt{1-v^2}}$; (2) analogously in the meridians on the $\phi_2-\phi_4$ plane, $\theta=\frac{\pi}{2}$, $\varphi=\frac{\pi}{2}$ or
$\frac{3\pi}{2}$, the kink solutions will be referred to as $K_2$ and are given by
$\psi_{K_2}(t,x)\, =\, 2\, {\rm arctan} \, e^{\pm
\sigma_2 \overline{x}}$ and (3) $K_3$ kinks, which live in the meridians on the $\phi_1-\phi_4$ plane, $\theta=\frac{\pi}{2}$, $\varphi=0$ or $\pi$, are $\psi_{K_3}(t,x) =\, 2\, {\rm arctan} \, e^{\pm \overline{x}}$. The topological ${\mathbb S}^3$-kink classical energies are: $E(K_1)= 2\lambda R^2 \sigma_3< E(K_2)= 2\lambda R^2 \sigma_2<
E(K_3)= 2\lambda R^2$.

Changing slightly the notation by denoting
$\psi=\theta^1$, $\theta=\theta^2$, $\varphi=\theta^3$, small
fluctuations around the kink solution
$\theta(x)=\theta_K(x)+\eta(x)=(\theta^1_K(x),\theta^2_K(x),
\theta^3_K(x))+(\eta^1(x),\eta^2(x),\eta^3(x))$ modify the action as:
\[
S[\theta^1,\theta^2,\theta^3]=S[\theta^1_K,\theta^2_K,\theta^3_K]+\frac{R^2}{2}\int
\, dtdx \, \eta(x)\Delta(K)\eta(x)+ {\cal O}(\eta^3) \qquad .
\]
The second-order operator governing the kink small fluctuations is
the geodesic deviation operator plus the Hessian of the
potential: $\Delta(K)\eta=-\left(
\nabla_{\theta_K^\prime}\nabla_{\theta_K^\prime}\eta+
R(\theta^\prime_K,\eta)\theta^\prime_K +\nabla_\eta {\rm
grad}V\right)$. Standard geometric calculations allow us to conclude that $K_1$
small fluctuations are governed by the matrix of Schr$\ddot{\rm
o}$dinger operators:
\begin{equation}
\Delta(K_1)=
\left(-\frac{d^2}{dx^2}-\frac{2\sigma^2}{{\rm cosh}^2 \sigma_3 x}\right){\mathbb I}+ {\rm diag}\left( \sigma_3^2, 1 ,\sigma_2^2\right)
\label{hess1}
\end{equation}
provided that a \lq\lq parallel
frame" to the kink orbit, i. e., fluctuations of the form
$\eta^2(x)={\rm cosh}\sigma_3 x \,\xi^2(x)$, $\eta^3(x)={\rm
cosh}\sigma_3 x \,\xi^3(x)$, is chosen.

Therefore, the meson spectrum in the $K_1$ kink sector has three
branches that share a perfectly transmitting P$\ddot{\rm
o}$sch-Teller well but have different thresholds. The first branch
corresponds to fluctuations tangent to the kink orbit. There is a
bound state, $\eta^1_0(x)=\frac{1}{{\rm cosh}\sigma_3x}$, of zero
eigenvalue and one-particle scattering states $\eta^1_k(x)=e^{i k
\sigma_3 x}({\rm tanh}\sigma_3 x - i k)$ with frequencies
$\omega^2(k)=\sigma_3^2(k^2+1)$. In the orthogonal directions the
eigenfunctions are the same but the bound state energies and
thresholds of the continuous spectra are shifted respectively to:
$1-\sigma_3^2$, $\sigma_2^2-\sigma_3^2$, $1$ and $\sigma_2^2$.

\section{Spectral zeta function and kink mass quantum correction}

We choose a normalization interval of length $l=\lambda L$ and
impose periodic boundary conditions on the fluctuations:
$\eta(-\frac{l}{2})=\eta(\frac{l}{2})$. At the end of the computations we will send the
length $l$ of the normalization interval to infinity. $\Delta(K)$ acts on the
Hilbert space $L^2=L^2_1({\mathbb S}^1)\bigoplus L^2_2({\mathbb
S}^1)\bigoplus L^2_3({\mathbb S}^1)$. The heat trace ($\beta$ is a fictitious inverse temperature or Euclidean time) is:
\[
{\rm Tr}_{L^2}e^{-\beta\Delta(K_1)}=
\frac{l {\cal A} }{\sqrt{4\pi\beta}} +
{\rm tanh}\frac{\sigma_3
l}{2}\left(1+e^{-(1-\sigma_3^2)\beta}+e^{-(\sigma_2^2-\sigma_3^2)\beta}\right){\rm
Erf}[\sigma_3\sqrt{\beta}]
\]
where ${\cal A} =e^{-\sigma_3^2\beta}+ e^{-\beta}+ e^{-\sigma_2^2\beta}$. It is interesting also to use the short time asymptotics of the heat
trace. Due to the structure of the second-order fluctuations operator (\ref{hess1}),
a power $\beta$ expansion of the heat trace is sensible\cite{AMAJM}{}:
\[
{\rm Tr}_{L^2}e^{-\beta\Delta(K_1)}={\rm
Tr}_{L^2}e^{-\beta\Delta_0(K_1)}\sum_{n=0}^\infty \,
c_n(K_1)\beta^n=\frac{{\cal A}}{\sqrt{4\pi\beta}}
\sum_{n=0}^\infty \, c_n(K_1)\beta^n \, ,
\]
where the coefficients are: $c_0(K_1)=l$, $c_n(K_1)=\frac{2^{n+1}\sigma_3^{2n-1}}{(2n-1)!!}$.

\noindent The Casimir energy $\Delta E^C=\Delta
E-\Delta E_0 =\frac{\lambda}{2}\left[{\rm
Tr}_{L^2}^*\Delta^\frac{1}{ 2}(K_1)-{\rm
Tr}_{L^2}\Delta_0^\frac{1}{2}(K_1)\right]$ is ultra-violet
divergent. We shall regularize these divergences by using the zeta function method. The zeta
functions are the Mellin transform of the heat traces, $\zeta_{\Delta}(s)= \frac{1}{\Gamma(s)} \int_0^\infty d\beta \beta^{s-1} \, {\rm Tr}_{L^2}\, e^{-\beta\Delta}$ and thus we regularize the divergence by assigning to it the
value of the spectral zeta function at a regular point of the
$s$-complex plane: $\Delta
E^C(s)=\frac{\mu}{2}\left(\frac{\mu^2}{\lambda^2}\right)^s
\left[\zeta_{\Delta(K_1)}^*(s)-\zeta_{\Delta_0(K_1)}(s)\right]$.
The behaviour of the kink
Casimir energy near the physical pole $s=-\frac{1}{2}+\varepsilon$
is:
\begin{equation}
\Delta E^C = -\frac{\lambda
\sigma_3}{2\pi} \left[ \frac{3}{\varepsilon} + 3 \ln
\frac{\mu^2}{\lambda^2} + \ln
\frac{2^6}{\sigma_3^2\bar{\sigma}_{13}^2\bar{\sigma}_{23}^2} -4 +
F[{\textstyle -\frac{\sigma_3^2}{\bar{\sigma}_{13}^2}}]+
F[{\textstyle -\frac{\sigma_3^2}{\bar{\sigma}_{23}^2}}]
\right] \label{kcep}
\end{equation}
where we denote $F[x]={}_2F_1^{(0,1,0,0)}[\frac{1}{2},0,\frac{3}{2};x]$,
$\bar{\sigma}_{13}^2=1-\sigma_3^2$ and $\bar{\sigma}_{23}^2=\sigma_2^2-\sigma_3^2$.

The kink energy due to the mass renormalization counter-terms that
must be added, $\Delta E^{MR}=-\lambda \frac{\sigma_3^2}{R^2} [I(1)+I(\sigma^2_2)+I(\sigma^2_3)] \int \, dx \, \phi_3^{{\rm K}_1}(x)\phi_3^{{\rm K}_1}(x)=2\lambda\sigma_3 [I(1)+I(\sigma_2^2)+I(\sigma_3^2)]$ is also ultra-violet divergent. The loop integrals become in the
finite length normalization interval divergent series susceptible of
being regularized as spectral zeta functions:
\[
I(c^2)=\frac{1}{2l}\sum_{n=-\infty}^\infty \, \frac{1}{(\sigma_3^2
n^2+c^2)^\frac{1}{2}}=-\frac{\lambda}{\mu l} \lim_{s\rightarrow
-\frac{1}{2}}
\left(\frac{\mu^2}{\lambda^2}\right)^{s+1} \frac{\Gamma(s+1)}{\Gamma(s)}\zeta_{-\frac{d^2}{dx^2}+c^2}(s)
\]
The regularized mass renormalization kink energy
\[
\Delta E^{\rm
MR}(s)=-\frac{2\sigma_3\lambda^2}{\mu\sqrt{4\pi}}
\left(\frac{\mu^2}{\lambda^2}\right)^{s+1}\frac{\Gamma(s+\frac{1}{2})}{\Gamma(s)}
\left(1+\frac{1}{\sigma_2^{2s+1}}+\frac{1}{\sigma_3^{2s+1}}\right)
\]
behaves near the physical pole as:
\begin{equation}
\Delta E^{\rm MR}(-\frac{1}{2}+\varepsilon)=  \frac{\lambda
\sigma_3}{2\pi} \left[ \frac{3}{\varepsilon}+ 3 \ln
\frac{\mu^2}{\lambda^2} +3(\ln 4-2)- \ln \sigma_2^2\sigma^2_3
\right]  \label{mrkep}
\end{equation}
From the short-time asymptotics of the heat trace we obtain an
approximated formula for the kink Casimir energy by means of the
partial Mellin transform on the $[0,b]$ integration interval of the
truncated to $N_0$ terms heat trace expansion:
\[
\Delta
E^C(b,N_0)=-\frac{\lambda}{2\sqrt{\pi b}}
-\frac{\lambda}{8 \pi} \sum_{n=1}^{N_0}\,
c_n(K)\left[\frac{\sigma_3^2}{\sigma_3^{2n}}\gamma[\sigma_3^2
b]+\gamma[b]+\frac{\sigma_2^2}{\sigma_2^{2n}}\gamma[\sigma_2^2
b]\right]
\]
where $\gamma[c]=\gamma[n-1,c]$ and $\gamma[z,c]$ is the incomplete Euler gamma
function. The contribution $\Delta E^C_{(1)}$ of the term with $c_1(K_1)=4\sigma_3$ to this approximation to the kink Casimir energy is divergent because $z=0$ is a pole of $\gamma[z,c]$. Fortunately, the divergent mass
renormalization kink energy $\Delta E^{\rm
MR}$ exactly cancels $\Delta E^C_{(1)}$.

Finally, the $K_1$ semiclassical mass, $E(K_1)= 2\lambda\sigma_3^2
R^2+\Delta E+ {\cal O}(\frac{1}{R^2})$, is obtained by adding
(\ref{kcep}) and (\ref{mrkep}):
\begin{equation}
\Delta E=
-\frac{\lambda\sigma_3}{2\pi}\left[2+ F[{\textstyle -\frac{\sigma_3^2}{\bar{\sigma}_{13}^2}}]
+ F[{\textstyle -\frac{\sigma_3^2}{\bar{\sigma}_{23}^2}}]+
\ln\frac{\sigma_2^2}{\bar{\sigma}_{13}^2\bar{\sigma}_{23}^2}\right] \, .
\label{olmsk11}
\end{equation}
Because the wells in the second-order fluctuation operator are
transparent the Cahill-Comtet-Glauber formula \cite{CCG}{}, $\Delta E(K_1)=-\frac{\lambda\sigma_3}{\pi}
[\sin\nu_1+\frac{1}{\sigma_3}\sin\nu_2+\frac{\sigma_2}{\sigma_3}\sin\nu_3-\nu_1\cos\nu_1
-\frac{1}{\sigma_3}\nu_2\cos\nu_2-\frac{\sigma_2}{\sigma_3}\nu_3\cos\nu_3]$, with
$\nu_1= \arccos(0)={\pi\over 2}$,  $\nu_2=\arccos \bar{\sigma}_{13}$
, $\nu_3=\arccos\frac{\bar{\sigma}_{23}}{\sigma_2}$, giving the
one-loop mass shift in terms only of the bound state eigenvalues and
the thresholds of the continuous spectra, can be
applied\cite{CCG}{}. Despite appearances, the result
\begin{equation} \Delta E(K_1) = -\frac{\lambda\sigma_3}{\pi}\left[
3-\frac{\bar{\sigma}_{13}}{\sigma_3}\arccos(\bar{\sigma}_{13})
-\frac{\bar{\sigma}_{23}}{\sigma_3}\arccos(\frac{\bar{\sigma}_{23}}{\sigma_2})
\right] \label{olmsk12}
\end{equation}
is identical to (\ref{olmsk11}) as one can check by plotting of
both expressions. A third (approximate) formula, useful in the cases
when the spectral information on the kink fluctuations is unknown, is derived from the asymptotic expansion:
\begin{equation}
\Delta
E(b,N_0)=-\frac{\lambda}{2\sqrt{\pi b}}
-\frac{\lambda}{8\pi}\sum_{n=2}^{N_0}\,
c_n(K)\left[\frac{\sigma_3^2}{\sigma_3^{2n}}\gamma[\sigma_3^2
b]+\gamma[b]+\frac{\sigma_2^2}{\sigma_2^{2n}}\gamma[\sigma_2^2
b]\right]\label{olmsk13}
\end{equation}

\bibliographystyle{ws-procs9x6}
\bibliography{ws-pro-sample}

\end{document}